\documentclass[12pt]{article}

\usepackage{amsfonts,amssymb}
\input epsf

\def\d{\mbox{d}}

\title{ \bf \Large
Towards precision determination of the top quark mass  from
$M_{b\ell}$ distribution in semi-leptonic decays}
\author{M. L. Nekrasov \\
{\small\it Institute for High Energy Physics, 142284 Protvino,
Russia}}

\date{}

\begin{document}

\maketitle

\begin{abstract}
We explore a possibility of optimization of the method of
determination of the top quark mass from $M_{b\ell}$ distribution
in semi-leptonic decays $t \to b \, \ell\nu$ at LHC and a future
linear collider (LC). We discover that the systematic and
statistical errors of $M_t$ determination can be diminished if
considering the high moments over the distribution. In the case of
LHC this allows one to reduce more than in twice the errors, and
in the case of LC to approach to the precision expected at
studying the threshold scan of the total cross-section $e^+e^- \to
t\bar t$.
\end{abstract}

\section{Introduction \label{Introduction}}

The precision determination of the top quark mass is one of the
major research problems at colliders of next generation
\cite{top,TESLA,Snowmass,ACFA,CLIC}. Being a fundamental parameter
of the Standard Model (SM), the top quark mass is tightly
constrained by quantum level calculations with other fundamental
parameters. This enables one to test the SM and/or to select the
probable scenario of its extension on the basis of an independent
$M_t$ measurement.

A considerable progress in this direction is expected at Tevatron
and LHC where the accuracy of the $M_t$ determination is
anticipated of about 1-2 GeV \cite{top}. At LHC in view of the
copious production of the top quarks, for the increase of the
accuracy the decrease of systematic errors is crucial. An analysis
of Ref.~\cite{top} shows that the method most promising from the
point of view of optimization of the errors is based on the
investigation of a distribution over the invariant mass of the
observable products of semi-leptonic decays $t \to b W\! \to b \,
\ell \nu \,$; more precisely of the isolated lepton $\ell$ and the
$\mu^{+}\mu^{-}$ pair indicating a $J/\psi$ meson produced from
the decay of the $b$ quark \cite{Kharchilava}. In this channel one
can obtain experimentally very clean final states.
Correspondingly, the systematic error of the $M_t$ measurement can
be made low. The evaluation made by Monte-Carlo (MC) modelling
gives 0.6-0.8 GeV at the statistical error of about 1 GeV for 4
years of LHC operation \cite{Kharchilava}. This result is
recognized as the best one among others obtained by various
methods \cite{top}.

In the case of a future linear collider (LC)
\cite{TESLA,Snowmass,ACFA,CLIC} the most promising method for
precision $M_t$ determination is based on the investigation of the
threshold scan of the total cross-section $e^+ e^- \to t\bar t$.
In this region the form and the height of the cross-section are
very sensitive to the mass of the top quark. This gives an
opportunity to determine $M_t$ with very high accuracy. A serious
difficulty in this approach is a precise theoretical calculation
of the behavior of the cross-section in the vicinity of the
threshold, which becomes additionally complicated because of the
resonant effects due to the strong $t$--$\bar t$ interaction. A
major progress in the calculations was made by way of the
summation of QCD contributions via solving Lippmann-Schwinger
equation for the Green function describing the $t \bar t$
production \cite{Khose}. At the present moment the theoretical
error of the top mass determined by this method is estimated at
100-200~MeV \cite{threshold,1S}, with the experimental error of
about 20 MeV~\cite{Martinez}.

Alternate methods of the $M_t$ determination are based on the
reconstruction of events of the decays of the top quarks. In the
basic features they are common at LC and at the hadron colliders,
but at LC the precision is anticipated better. Thus, for example,
the systematic error of $M_t$ determination by direct
reconstruction of $t \bar t$ events in $e^+ e^-$ collisions at
$\sqrt{s} = 500$ GeV is expected \cite{Chekanov1} at 340 and 250
MeV in hadronic and semi-leptonic channels, respectively, with
statistical errors of about 100 MeV for 1-2 years of LC
\cite{Chekanov2}. Since far above the threshold one can expect
very high precision of the necessary theoretical calculations, the
resultant errors should be close to that expected at studying the
threshold scan of the cross-section. This promising anticipation
again excites a question about the precision of the top mass
determination by the method of Ref.~\cite{Kharchilava}, but this
time in the LC case. Actually this method in the LC case has been
discussed as a preliminary in \cite{Corcella1} (see also review
\cite{Snowmass}), but the errors have not been determined. So the
prospect of this method at LC is still not practically known.

In this article we clear up this question. In contrast to
Ref.~\cite{Kharchilava}, however, we consider the full
reconstructed jet of the $b$ quark instead of $J/\psi$ or
$\mu^{+}\mu^{-}$ pair only. Such an approach has been considered
in \cite{Corcella1}, and partially in \cite{Corcella2}. We follow
it by keeping in mind that the $M_{b\ell}$ distribution in any
case does emerge in a certain stage of the analysis. So from very
beginning the analysis can be made in terms of the data converted
to the form of $M_{b\ell}$ distribution. (Of course, the
systematic errors that arise in the course of the converting of
the data must be taken into account.) An obvious advantage of this
approach is a possibility to consider the data in a uniform
fashion in both cases, LHC and LC. Moreover, this allow us in a
simple way to explore a possibility of optimization of the
algorithm of the extraction of the top mass from the data. The
elaboration of the latter problem actually is the major purpose of
the present article.

In the next section we detail the statement of the problem. In
sections \ref{Stat} and \ref{Sys} we discuss a model for the
calculation of the errors. The parameters of the model are fixed
in section \ref{Results} and in the same place the quantitative
outcomes are determined. In section \ref{theor} we discuss the
theoretical uncertainty, and in section \ref{Discussion} we
discuss the~results.

\section{Statement of the problem\label{Statement}}

We consider the processes
\begin{equation}\label{Eq1}
e^+ e^- \, (q \bar q,\, gg) \to t \bar t \to b W \, b W \to b \ell
\nu \,\, b  q_1 q_2 \to \{ b\mbox{-jet} + \ell\} + \{3 \mbox{
jets}\}\,,
\end{equation}
with the $b$-jet, isolated lepton $\ell = \{e,\mu\}$, and
invisible in the final states neutrino coming from one $t$ quark,
and the remaining three jets coming from another $t$ quark. In the
experiment the above mentioned states are registered, and measured
is a distribution
\begin{equation}\label{Eq2}
F(q) =\frac{1}{\sigma} \, \frac{\d \sigma}{\d q}\,.
\end{equation}
Here $\sigma$ is the cross-section of the process (\ref{Eq1}), $q
\equiv M_{b\ell}$ is the reconstructed invariant mass of the
system $\{b\mbox{-jet} + \ell\}$.

We simulate the results of the experiment under the following
suppositions. First we suppose that there is a satisfactory method
for extracting signal from the data. Actually this means the
existence of a satisfactory model for the background processes
that survive after setting of kinematic cuts.\footnote{The set of
the cuts and the background processes in the LHC case have been
discussed in Ref.~\cite{Kharchilava}. In the LC case that has been
done in Refs.~\cite{Chekanov1,Chekanov2}. At this stage we do not
take manifestly into account the kinematic cuts but do that at
deriving the quantitative outcomes.} Further, we describe the
signal in the Born approximation, identifying the $b\mbox{-jet}$
with the $b$ quark. Finally, on the basis of the results of
Ref.~\cite{Kharchilava} we disregard the effects of finite width
of the top quarks. The latter assumption means that $\sigma^{-1}
\, \d\sigma/\d q$ is equal to $\Gamma^{-1}_{b \ell \nu}\,\d
\Gamma_{b \ell \nu} / \d q$, where $\Gamma_{b \ell \nu}$ is a
partial width of the decay $t \to b\ell\nu$. (Thus the
distribution $F$ becomes process-independent.)

\begin{figure}
\hbox{ \hspace*{-5pt}
       \epsfxsize=200pt \epsfbox{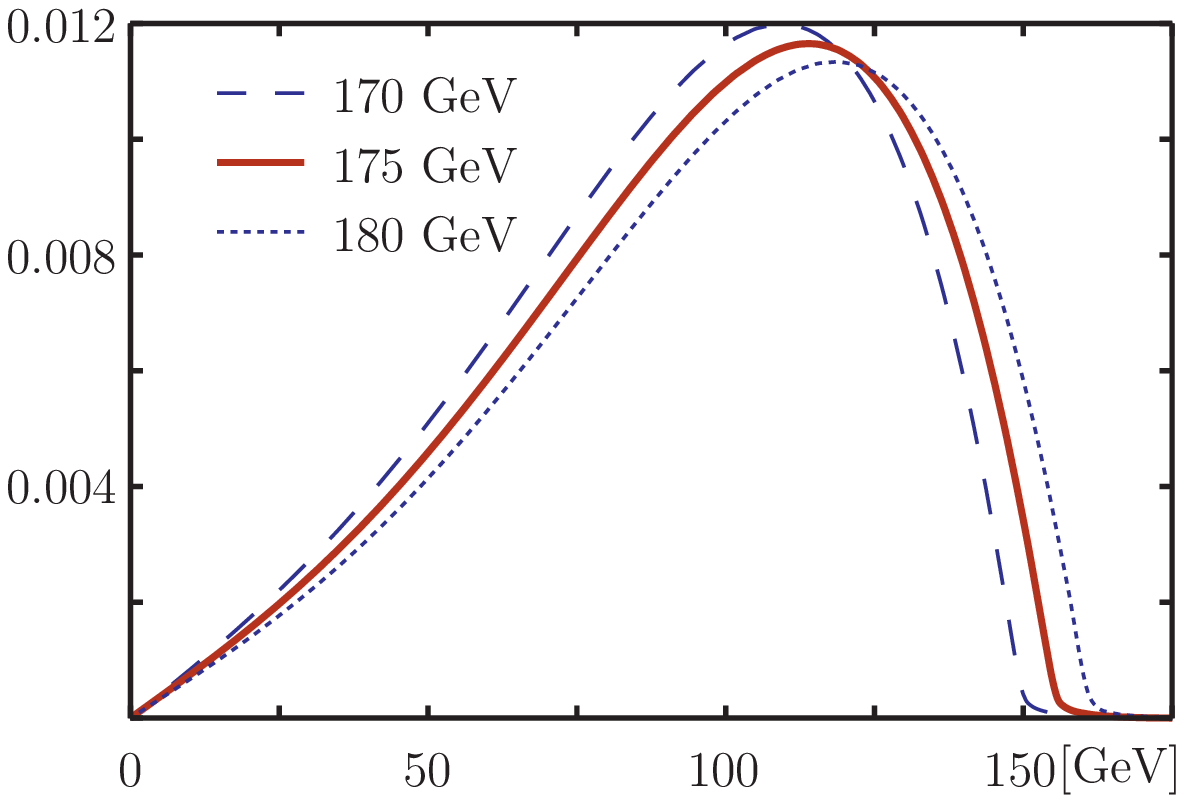}}
\caption{\small Distribution $F(q)\!=\!\Gamma^{-1} \, \d \Gamma /
\d q$, $q \equiv M_{b\ell}$, at $M_t =$ 170, 175, and 180 GeV.
}\label{Fig1}
\end{figure}

Direct calculation gives the following formula for the
distribution of the partial width:
\begin{eqnarray}\label{Eq3}
\lefteqn{\frac{\d \Gamma_{b \ell \nu}}{\d q^2} = \frac{3\,G_F
|V_{tb}|^2}{4 \sqrt{2} \, \pi^2} \,
\frac{\Gamma_{W\!\to\ell\nu}\,M_W}{M_t^3} }
\\
&& \times \> \Biggl\{q^2-\Lambda^2-M_W^2 \Biggr. +
\left(\frac{\Lambda^2\!-\! M_W^2}{2}-q^2\right) \ln \!
\frac{(\Lambda^2 - q^2)^2 + M_W^2 \Gamma_W^2}{M_W^4 + M_W^2
\Gamma_W^2} \nonumber \\
&& \qquad + \left. \frac{(\Lambda^2-q^2)(q^2+M_W^2) + M_W^2
\Gamma_W^2}{M_W \Gamma_W}\left[\arctan \!\! \left(\frac{\Lambda^2
- q^2}{M_W\Gamma_W}\right)+\arctan \!\! \left(
\frac{M_W}{\Gamma_W}\right)\right]\right\}\,.\nonumber
\end{eqnarray}
Here $\Lambda^2 = M_t^2-M_W^2$, $\Gamma_{W}$ is the total and
$\Gamma_{W\!\to\ell\nu}$ is the partial width of the $W$ boson,
and we neglect the masses of the lepton $\ell$ and the $b$
quark.\footnote{The influence of the mass of the $b$ quark is
noticeable at very small $q$, but this region is inessential when
considering the moments over the distribution.} In this
approximation $\Gamma_{W\!\to\ell\nu} = 2/9\, \Gamma_{W}$ and $q$
ranges between 0 and $M_t$. Fig.~\ref{Fig1} shows the distribution
$F(q)$ defined by formula (\ref{Eq3}) at $M_t = $ 170, 175, 180
GeV. From Fig.~\ref{Fig1} a dependence of $F(q)$ on $M_t$ is
obvious. So, by comparing the experimental distribution with a set
of theoretical curves one can determine, in principle, the
experimental value of $M_t$.

In a practical respect, however, it is convenient to compare
integrated parameters of the distributions. For instance, in
Ref.~\cite{Kharchilava} the $M_t$ was extracted from the mean
value (position of the maximum) of the Gaussian distribution
approximating the measured distribution.
Refs.~\cite{Corcella1,Corcella2} determined $M_t$ by the first
moment $\langle \, q \, \rangle$ over the distribution. In the
present article we consider a method of $M_t$ determination by the
higher moments
\begin{equation}\label{Eq4}
\langle q^n \rangle = \int_0^{M} \!\!\! \d q \; q^n F(q)\,.
\end{equation}
Here $M$ is a fixed quantity close to $M_t$, see below for
details. In fact this method means the matching of the
experimental distribution $q^n F(q)$ with the corresponding
theoretical distribution which depends on the parameter~$M_t$.

As we will see below, the insertion of $q^n$ factor will
significantly increase the precision of the $M_t$ determination.
Eventually this can be checked by a quantitative analysis.
Nevertheless some hints on this can be a priori seen. Really, with
increasing $n$ the moment $\langle q^n \rangle$ becomes in rising
measure dependent on the behavior of $F(q)$ in a region located
between the position of its maximum and a large-$q$ tail where
$F(q)$ almost vanishes. (More precisely, by the tail we mean a
range $\Lambda < q < M_t$, where in the limit $\Gamma_W=0$ the
distribution identically vanishes by reason of kinematics.)
Further, in the mentioned region the behavior of $F(q)$ in the
greatest measure is sensitive to the value of $M_t$, which is seen
from Fig.~\ref{Fig1}. As a result, with increasing $n$ the
sensitivity of $\langle q^n \rangle$ with respect to $M_t$ is
increasing. That is why one can expect the increasing of the
precision of $M_t$ extracted from the higher moments.

Now let us dwell on the details of definition (\ref{Eq4}). The
point of the discussion is the upper limit in the integral. We set
it $M$ instead of conventional $M_t$, meaning the upper bound of
the region allowed by kinematics, because $M_t$ also is a
parameter which is subject to determination. In order to avoid an
inconvenience, we put in the place of the upper limit a certain
predetermined value $M$ fixed close to $M_t$. Simultaneously we
adjust the normalization of the distribution $F(q)$ so that to
provide the equality $\langle 1 \rangle =1$. The moments $\langle
q^n \rangle$ at $n \ge 1$ after this redefinition practically do
not change (at not too large $n$) in view of almost vanishing $F$
in the tail at large $q$.

So, we define the experimentally measured value of the top quark
mass as a solution to the equation
\begin{equation}\label{Eq5}
\langle q^n \rangle = \langle q^n \rangle_{\mbox{\scriptsize
exp}}\,.
\end{equation}
Here in the r.h.s.~the moment is determined (at a given $M$) on
the basis of the experimental data, and that in the l.h.s.~on the
basis of the theoretical distribution which depends on the
parameter $M_t$. Let at a given $n$ a solution to the equation
(\ref{Eq5}) be $M_t = M_{t(n)}$. Then the error of the solution
can be determined as
\begin{equation}\label{Eq6}
\Delta \! M_{t(n)} = \Delta \langle q^n \rangle_{\mbox{\scriptsize
exp}}\!\left/ \frac{\d \langle q^n \rangle}{\d
M_t}\Biggl|_{M_t=M_{t(n)}}\Biggr. \right..
\end{equation}
Our aim is to estimate $\Delta \! M_{t(n)}$ and find an optimal
value of $n$ which would minimize $\Delta \! M_{t(n)}$. Since by
virtue of (\ref{Eq3}) the derivative $\d\langle q^n \rangle/\d
M_t$ is known, the problem is reduced to the determination of the
statistical and systematic errors, the components of the
experimental error $\Delta \langle q^n \rangle_{\mbox{\scriptsize
exp}}$.

\section{Statistical errors \label{Stat}}

We determine the statistical errors of the moments on the
supposition that the data averaged over ensemble are described by
$F(q) = \Gamma^{-1}_{b \ell \nu} \, \d \Gamma_{b \ell \nu} / \d q$
with $\Gamma_{b \ell \nu}$ determined by formula (\ref{Eq3}) at
$M_t=175$ GeV.

Let $\delta q_i$ be the size of a bin, within which $i$-th element
of the distribution is measured, and let $\overline{N_i}$ be the
number of events counted in this bin on the average. Then
\begin{equation}\label{Eq7}
F(q_i) \delta q_i = \overline{N_i}\,/\,\overline{N}\,.
\end{equation}
Here $\overline{N}$ is the total number of events counted in all
bins on the average. Further we do not distinguish between
$\overline{N}$ and $N = \sum_i N_i$, the total number of events
counted in all bins in the given experiment. The experimentally
measured $n$-th moment is
\begin{equation}\label{Eq8}
\langle q^n \rangle_{\mbox{\scriptsize exp}} = \sum_i q_i^n \,
\frac{N_i}{N}\,.\\[-0.4\baselineskip]
\end{equation}
By virtue of (\ref{Eq7}) the averaged experimental moment
$\overline{\langle q^n \rangle}_{\mbox{\scriptsize exp}}$ is found
by formula (\ref{Eq4}). Since $N_i$ is distributed by Poisson law
with parameter $\overline{N_i}$, the variance of $\langle q^n
\rangle_{\mbox{\scriptsize exp}}$ is
\begin{equation}\label{Eq9}
D\langle q^n \rangle_{\mbox{\scriptsize exp}} = \sum_i q_i^{2n} \,
\frac{\overline{N_i}}{N^2} \equiv \frac{1}{N}\langle q^{2n}
\rangle\,.
\end{equation}
Formula (\ref{Eq9}) implies the following estimation for the
statistical error:
\begin{equation}\label{Eq10}
\Delta^{\mbox{\scriptsize stat}}\langle q^n
\rangle_{\mbox{\scriptsize exp}} = \sqrt{\frac{1}{N}\langle q^{2n}
\rangle}\,.
\end{equation}

\begin{figure}
\hbox{ \hspace*{3pt}
       \epsfxsize=192pt \epsfbox{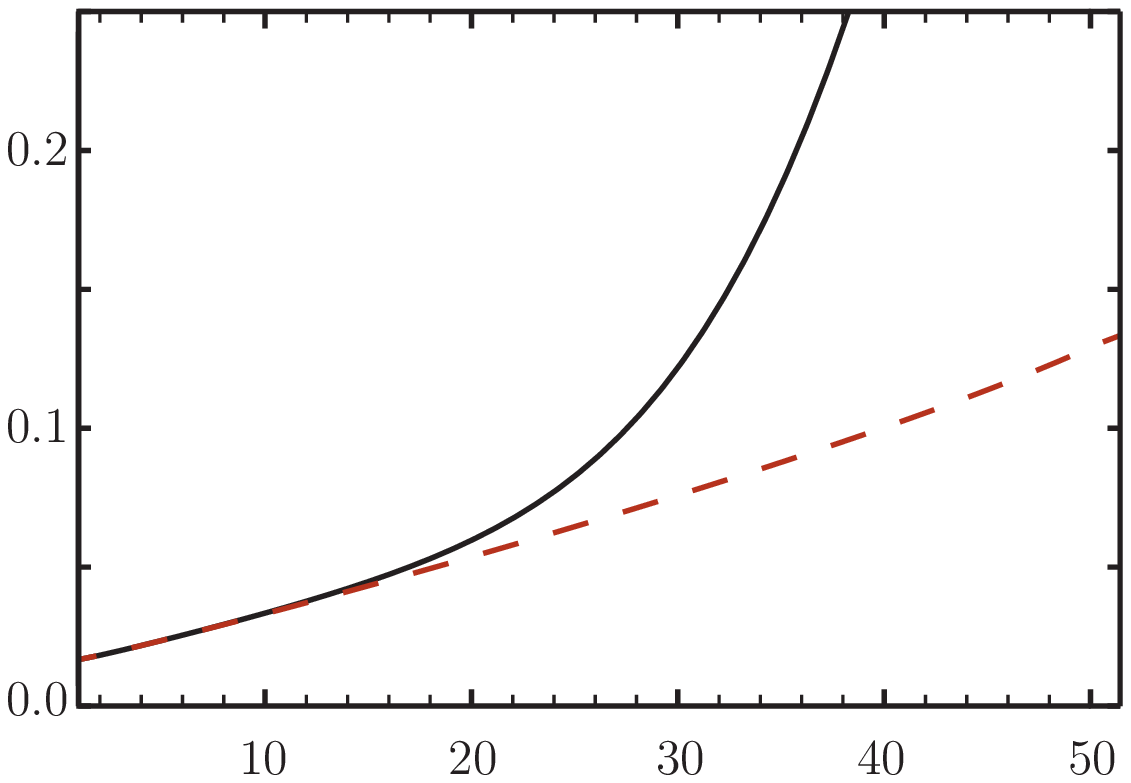}}
\caption{\small The ratio $\Delta^{\mbox{\scriptsize stat}}\langle
q^n \rangle_{\mbox{\scriptsize exp}}/\langle q^n
\rangle_{\mbox{\scriptsize exp}}$ depending on $n$ ($M_t\! =\!
175$ GeV, $N\!=\!4000$). The continuous curve represents the
results described by formula (\ref{Eq10}). The dashed curve
represents the results obtained by the method of the effective
moments.}\label{Fig2}
\end{figure}

\begin{figure}
\hbox{ \hspace*{10pt}
       \epsfxsize=192pt \epsfbox{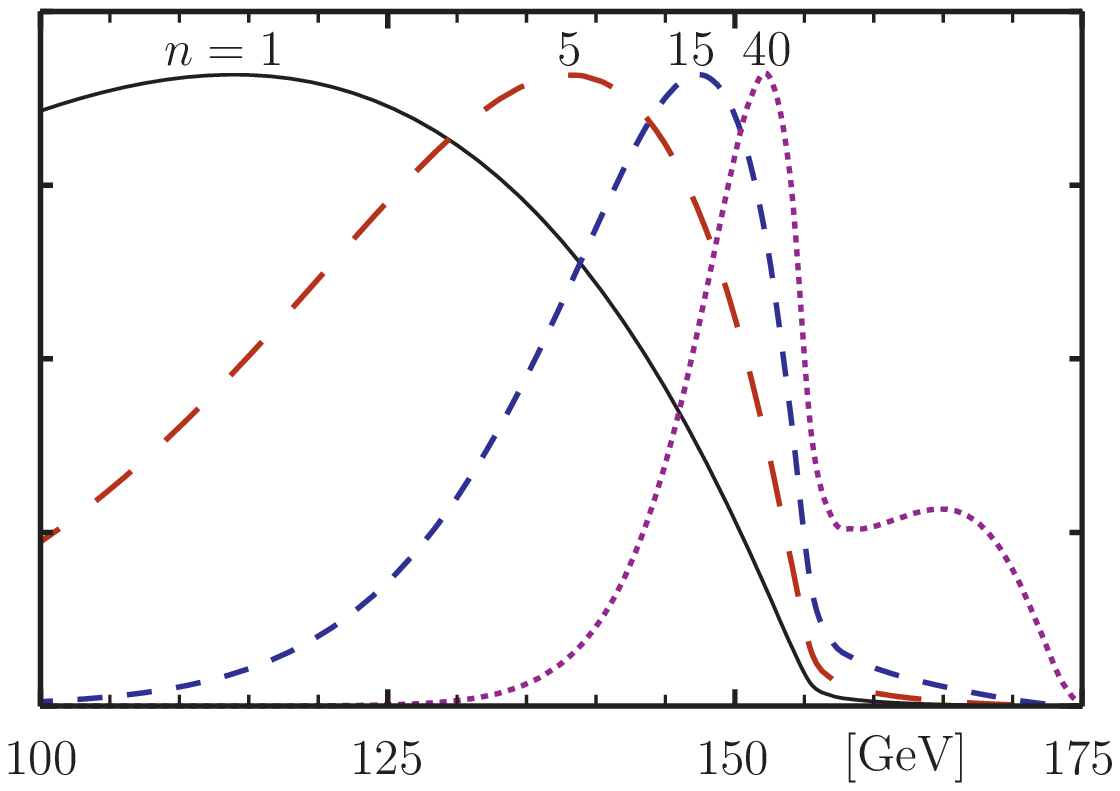}}
\caption{\small The shape of the function $q^n F(q)$ at $M_t =$
175 GeV, $n=1$, 5, 15, 40 (in arbitrary
normalization).}\label{Fig3}
\end{figure}

To give an idea of the behavior of $\Delta^{\mbox{\scriptsize
stat}}\langle q^n \rangle_{\mbox{\scriptsize exp}}$, we pre\-sent
in Fig.~\ref{Fig2} by a continuous curve the ratio
$\Delta^{\mbox{\scriptsize stat}}\langle q^n
\rangle_{\mbox{\scriptsize exp}}$/ $\langle q^n
\rangle_{\mbox{\scriptsize exp}}$ calculated at $N=4000$
(corresponds to LHC case, see Sect.~\ref{Results}). It is seen
from the figure that with increasing $n$ the ratio is growing.
This is explained by the shift (to the right) of the position of
maximum of $q^n F(q)$ from the position of maximum of $F(q)$,
where the statistics is largest. As a result a statistical
reliability of $\langle q^n \rangle_{\mbox{\scriptsize exp}}$
comes down. Another important property of the ratio is the change
of the mode of the growth beginning with $n \approx 15 $. This is
explained by the emergence of a noticeable contribution from the
large-$q$ tail in $q^n F(q)$. The latter property is illustrated
by the set of the curves represented by Fig.~\ref{Fig3}.

In fact the emergence of a noticeable contribution from the tail
is an undesirable effect since on the tail the uncertainty from
the background is comparable with the signal process. In order to
avoid this difficulty one can correct the definition of the
moments by introducing a cutoff in the integral in (\ref{Eq4}).
The position of the cutoff should be determined so that to isolate
the second (unphysical) peak in the tail of $q^n F(q)$ but
simultaneously to keep as much as possible a statistical
significance of the sample of events. It is clear that the optimal
cutoff should be placed in the neighborhood of a local minimum
between the two peaks of $q^n F(q)$ (if the second peak appears).
From Fig.~\ref{Fig3} it is seen that at $n \approx 40 $ the
sought-for point is distant by about two half-widths to the right
of the position of the maximum of $q^n F(q)$. So a simplified
algorithm for the cutoff may be determined by setting $\Lambda_{n}
= \min\{q_{n\,\mbox{\scriptsize extr}} + 2
\Gamma_{n\,\mbox{\scriptsize right}},M\}$, where
$q_{n\,\mbox{\scriptsize extr}}$ is the position of the maximum of
$q^n F(q)$ and $\Gamma_{n\,\mbox{\scriptsize right}}$ is the
half-width from the right. Thus we come to the following
definition of the effective moments:
\begin{equation}\label{Eq11}
\langle q^n \rangle^{\mbox{\scriptsize eff}} = \int_{0}^{
\Lambda_{ \mbox{\scriptsize n}}} \!\!\! \d q \; q^n F(q)\left/
\int_{0}^{ \Lambda_{ \mbox{\scriptsize n}}} \!\!\! \d q \; F(q)
\right. .
\end{equation}

In the experimentally determined effective moments the cutoff must
be the same. Ultimately $\Delta^{\mbox{\scriptsize stat}}\langle
q^n \rangle_{\mbox{\scriptsize exp}}^{\mbox{\scriptsize eff}}$ is
defined by formula (\ref{Eq10}) with $\langle q^{2n} \rangle$
replaced by $\langle q^{2n} \rangle^{\mbox{\scriptsize eff}}$ but
with introducing the cutoff $\Lambda_{n}$ instead of
$\Lambda_{2n}$. The latter anomalous prescription follows
immediately from the derivation of formula (\ref{Eq10}).

The behavior of $\Delta^{\mbox{\scriptsize stat}}\langle q^n
\rangle_{\mbox{\scriptsize exp}}^{\mbox{\scriptsize eff}}/\langle
q^n \rangle_{\mbox{\scriptsize exp}}^{\mbox{\scriptsize eff}}$ is
shown by the dashed curve in Fig.~\ref{Fig2}. It is seen from the
figure that the transition to the effective moments implies no
noticeable modification up to $n \approx 15$, while at the larger
$n$ the growth of the ratio becomes stabilized. A similar behavior
is observed in the basic formalism (without the transition to the
effective moments) in the limit $\Gamma_{W} \to 0$, when the
large-$q$ tail identically vanishes.

\section{Systematic errors \label{Sys}}

Proceeding to the systematic errors it is necessary at first to
clarify a reason of their origin. For this purpose we use the
analysis of Ref.~\cite{Kharchilava} of the errors of the
$M_{\mbox{\scriptsize \rm{J}}\!/\!\psi\ell}$ distribution
simulated with the PYTHIA and/or HERVIG event generators. By the
main sources of the systematic errors Ref.~\cite{Kharchilava}
found the uncertainties in the $b$ quark fragmentation (including
the final state radiation) and the uncertainties in the background
processes. It is clear that the same sources should be the main
ones at solving the inverse problem, the determination of the
$M_{b\ell}$ distribution from
$M_{\mbox{\scriptsize\rm{J}}\!/\!\psi\ell}$ distribution which is
considered virtually as the data. In the LC case we expect the
same pattern of the origin of systematic errors.

On this basis we consider at first the error resulting from the
uncertainty in the $b$ quark fragmentation. For brevity we call it
by the error of the type I. At the level of $M_{b\ell}$
distribution it appears as the uncertainty in the number of the
bin within which the number of events, $N_i$, is measured. In the
continuous case this error becomes the uncertainty $\Delta q$ in
the determining of $q$ variable.

Suppose that $\Delta q$ is sufficiently small. Then, neglecting
the nonlinear effects, we have
\begin{equation}\label{Eq12}
\Delta^{\mbox{\scriptsize sys I}}\langle q^n
\rangle_{\mbox{\scriptsize exp}} = \int_0^{M} \!\!\! \d q \; \,
\left[q^n  F(q)\right]^{\prime}\, \Delta q \,.
\end{equation}
Here the prime means the derivative with respect to $q$. The
systematic error I of the effective moment $\langle q^n
\rangle_{\mbox{\scriptsize exp}}^{\mbox{\scriptsize eff}}$ is
estimated similarly, with replacing the upper bound $M$ by
$\Lambda_{n}$ and, then, dividing the result by the normalization
factor as in formula (\ref{Eq11}). The normalization factor itself
should be the same as it controls the total number of events that
are taken into consideration at the determining of the effective
moment.

The determination of $\Delta q$ we carry out with the aid of the
following reasoning. First we note that the invariant mass $q^2$
actually is the doubled scalar product of 4-momenta of the $b$
quark and the lepton $\ell$. So in the laboratory frame it can be
represented as $q^2 = E_b \, K$, where $E_b$ is the energy of the
$b$ quark, and $K$ is a factor proportional to the energy of the
lepton $\ell$. (Additionally $K$ includes a dependence on angular
variables which, however, is relatively weak.) Further, by the
calculating of the differential we get $\Delta q =
\frac{1}{2}\left(\Delta E_b/E_b + \Delta K/K\right)q$, where
$\Delta E_b$ and $\Delta K$ are the corresponding errors. A more
precise estimation is determined by the sum in the quadratures.
Thus we come to a linear dependence with a certain coefficient
$r$,
\begin{equation}\label{Eq13}
\Delta q = r \, q \,,\qquad r=\frac{1}{2}\left[\left(\frac{\Delta
E_b}{E_b}\right)^2 + \left(\frac{\Delta
K}{K}\right)^2\right]^{1/2}.
\end{equation}

The systematic error arising after subtraction of the background
we name by the error of type II. It appears in the absolute value
of the distribution function. So it should be described as an
additive contribution $\delta F$ to function $F$. Correspondingly,
we get the following formula for the error II of the moments:
\begin{equation}\label{Eq14}
\Delta^{\mbox{\scriptsize sys II}}\langle q^n
\rangle_{\mbox{\scriptsize exp}} = \int_0^{M} \!\!\! \d q \; \,
q^n \, \delta F(q) \,.
\end{equation}
The error II of the effective moments $\langle q^n
\rangle_{\mbox{\scriptsize exp}}^{\mbox{\scriptsize eff}}$ is
defined by a similar formula to within modifications listed below
(\ref{Eq12}).

It is reasonable to determine $\delta F(q)$ on supposition that it
vanishes at the boundaries of the phase space and when passing
from small $q$ to large $q$ it only once changes the sign. The
simplest form of a function satisfying to these requirements is a
polynomial of degree three,
\begin{equation}\label{Eq15}
\delta F = h \, q\,(q-M/2)\,(q-M).
\end{equation}
Parameter $h$ in (\ref{Eq15}) describes the amplitude of the error
and it is subject to further determination.

\section{Numerical results\label{Results}}

We assign the following values for the parameters having a global
meaning:
\begin{equation}\label{Eq16}
M_W=80.4 \mbox{ GeV}, \; \Gamma_W=2.1 \mbox{ GeV}, \; M_t \! = \!
M \! = \! 175 \mbox{ GeV}\,.
\end{equation}
The remaining parameters $N$, $r$, and $h$ depend on the
conditions of the consideration. Recall that $N$ means the volume
of the representative sample of events, parameter $r$
characterizes the error in the invariant mass of the
$b\ell$-system, and $h$ describes the error arising after the
subtraction of the background processes.


With reference to LHC case, the parameters $N$, $r$,~$h$ we fixe
on the basis of the results of Ref.~\cite{Kharchilava}. Since in
that work the $M_{\mbox{\scriptsize \rm{J}}\! / \!\psi\ell}$
distribution was determined at $N = 4000$ (with kinematic cuts and
for 4 years of LHC), in our investigation we set this value for
$N$, as well. Parameters $r$ and $h$ we fix based on the
properties of $M_{\mbox{\scriptsize \rm{J}} \! / \!\psi\ell}$
distribution and the direct results derived in
Ref.~\cite{Kharchilava} from these properties. First we use the
estimation $\Delta^{\mbox{\scriptsize sys}}\langle
M_{\mbox{\scriptsize \rm{J}}\!/\!\psi\ell} \rangle =
+0.3/\!\!-\!0.4$ GeV and the derived from it the result $\Delta
M_t=+0.6/\!\!-\!0.8$~GeV. Considering in the framework of our
investigation the latter quantity as the uncertainty of the input
parameter $M_t$, we get by direct calculation
$\Delta^{\mbox{\scriptsize sys}}\langle q
\rangle_{\mbox{\scriptsize exp}} = +0.47/\!\!-\!0.62$ GeV. By
comparing this with $\Delta^{\mbox{\scriptsize sys}}\langle
M_{\mbox{\scriptsize \rm{J}}\!/\!\psi\ell} \rangle$ we obtain an
energy scale factor of 1.6, which describes the spreading of the
$M_{\mbox{\scriptsize \rm{J}}\!/\!\psi\ell}$ distribution when
converting it to the $M_{b\ell}$ distribution. Using the mentioned
factor, from $\Delta^{\mbox{\scriptsize sys II}}\langle
M_{\mbox{\scriptsize \rm{J}}\!/\!\psi\ell}
\rangle_{\mbox{\scriptsize exp}}$ {\small $\lesssim$} 0.15 GeV
\cite{Kharchilava} we further derive an estimation
$\Delta^{\mbox{\scriptsize sys II}}\langle q
\rangle_{\mbox{\scriptsize exp}}$ {\small $\lesssim$} 0.24 GeV.
From this result and (\ref{Eq14}), (\ref{Eq15}) we get $h \simeq
1.7 \times 10^{-10}$ GeV$^{-4}$. (Hereinafter we take the upper
bounds as the estimations.)

Knowing $\Delta^{\mbox{\scriptsize sys}}\langle q
\rangle_{\mbox{\scriptsize exp}}$ and $\Delta^{\mbox{\scriptsize
sys II}}\langle q \rangle_{\mbox{\scriptsize exp}}$, we
immediately get $\Delta^{\mbox{\scriptsize sys I}}\langle q
\rangle_{\mbox{\scriptsize exp}} \simeq +0.41 / \!\!-\!0.57$. From
here and formula (\ref{Eq12}) there follows $r \simeq$
0.004-0.006. Further we use the average value $r=0.005$. It is
worth noticing that the same estimation for $r$ follows from
formula (\ref{Eq13}) when taking into consideration the
1\%-precision of the determination of the energy of $b$ jets
expected at LHC \cite{top}, and additionally neglecting $\Delta
K/K$ as compared to $\Delta E_b/E_b$.

Now as we know $N$, $r$, $h$, we can calculate
$\Delta^{\mbox{\scriptsize stat}} \langle q^n
\rangle_{\mbox{\scriptsize exp}}$ and $\Delta^{\mbox{\scriptsize
sys I,II}} \langle q^n \rangle_{\mbox{\scriptsize exp}}$ at any
$n$. Then we calculate $\Delta^{\mbox{\scriptsize stat}} M_{t(n)}$
and $\Delta^{\mbox{\scriptsize sys I,II}} M_{t(n)}$. The
dependence on $n$ of these errors is shown by the solid lines on
Figs.~\ref{Fig4}--\ref{Fig6}. The dashed lines show the errors
obtained by the method of the effective moments. (The break of
slope in the dashed line in Fig.~5 is explained by the change of
the sign in the integral in formula (\ref{Eq12}) appearing after
introducing the cutoff $\Lambda_n$.) In Table~\ref{T1} we present
the numerical results at some $n$. In the same place we show the
summed in the quadrature errors $\Delta^{\mbox{\scriptsize sys}}
M_{t(n)}$ and $\Delta \! M_{t(n)}$. It should be noted that at
$n=1$ the systematic errors in Table~\ref{T1} practically coincide
with those in \cite{Kharchilava}. The reason is that we have fixed
the parameters of the model actually by matching the errors of the
first moments.

\begin{table}[t]
  \centering
\begin{tabular}{c c c c c c }
\hline
& & & & & \\[-3mm]
  n & $\Delta^{\mbox{\scriptsize stat}} M_{t(n)}$
  & $\Delta^{\mbox{\scriptsize sys I} } M_{t(n)}$
  & $\Delta^{\mbox{\scriptsize sys II}} M_{t(n)}$
  & $\Delta^{\mbox{\scriptsize sys}   } M_{t(n)}$
  & $\Delta \! M_{t(n)}$ \\[1mm] \hline
& & & & & \\[-3mm]
 1  & 2.07       & 0.62        & 0.30        & 0.69
                                                   & 2.18 \\
 5  &0.62        & 0.13        & 0.20 (0.17) & 0.24 (0.21)
                                                   & 0.66 \\
 10 &0.45        & 0.07 (0.06) & 0.20 (0.13) & 0.21 (0.14)
                                                   & 0.50 (0.48) \\
 15 &0.41 (0.40) & 0.05 (0.03) & 0.24 (0.12) & 0.24 (0.12)
                                                   & 0.48 (0.42) \\
 20 &0.42 (0.39) & 0.03 (0.00) & 0.32 (0.11) & 0.32 (0.11)
                                                   & 0.52 (0.40) \\
 30 &0.59 (0.38) & 0.02 (0.03) & 0.63 (0.11) & 0.63 (0.11)
                                                   & 0.86 (0.39) \\
 [0.5mm] \hline
\end{tabular}
\caption{\small Statistical, systematic I and II, and the
systematic summed in quadrature errors presented in GeV in the LHC
case. The last column represents the sum of the statistical and
the systematic errors. In the brackets we show the results
calculated by the method of the effective moments (if they are
different with the results calculated by the basic
method).}\label{T1}
\end{table}

In the LC case, unfortunately, there are no published results that
could allow us in a similar way to fix the parameters of the
model. Therefore we make use mainly of indirect methods. Parameter
$N$ we fix by the following reasoning. First we note that
$\sigma(e^+ e^- \to t\bar t) \approx 0.6$~pb at $\sqrt{s} = 500$
GeV. So at the integrated luminosity of 300 fb$^{-1}$,
corresponding to 1-2 years of running, approximately $180 \, 000$
$t\bar t$ pairs must be generated. Since the branching of the
process (\ref{Eq1}) is near of 30\%, only $54\,000$ events of
$t\bar t$ are related to our investigation. The efficiency of
their detection we estimate as follows. Suppose that at LC the
efficiency of the detecting of $W$-pairs decaying in a
semi-leptonic channel will be the same as at LEP2, i.e. $\sim
80$\% \cite{LEP2}. In addition, following Ref.~\cite{Snowmass}, we
suppose that the $b$-jet tagging efficiency at LC will be about
80\%. In summary this gives an acceptance of 50\% which implies $N
= 27\,000$.

Parameter $r$ we fix based on the systematic error of $M_t$
obtained in \cite{Chekanov1} in the approach of the direct
reconstruction of $t \bar t$ events in the semi-leptonic channel.
Additionally we use the note that in the kinematic range near the
upper $M_{b\ell}$ endpoint the neutrino practically does not
contribute to the total invariant mass of the decay products of
the top quark. Therefore the determination of the $M_t$ in the
mentioned range practically is the same that the determination of
the $M_{b\ell}$ invariant mass. Ref.~\cite{Chekanov1} obtained
$\Delta^{\mbox{\scriptsize sys}} M_t = 250$ MeV. So we set
$\Delta^{\mbox{\scriptsize sys}} M_{b\ell} = 250$ MeV. Assuming
that this is the error of the type I, we equate $\Delta q$ to this
value. Finally, by setting $\Delta q = rq$, $q \simeq M_t$ we get
$r = 0.0014$.

Parameter $h$ we fix by proceeding to the note of
Ref.~\cite{Kharchilava} about the decreasing of the systematic
error II of the average $\langle M_{\mbox{\scriptsize
\rm{J}}\!/\!\psi\ell} \rangle$ below of 0.1 GeV at the increasing
of statistics up to $N \sim 10^4$. From this by using the above
method we get a rough estimate $h \simeq 1.1 \times 10^{-10}$
GeV$^{-4}$.

Knowing $N$, $r$, $h$, we find $\Delta^{\mbox{\scriptsize stat}}
\langle q^n \rangle_{\mbox{\scriptsize exp}}$,
$\Delta^{\mbox{\scriptsize sys I,II}} \langle q^n
\rangle_{\mbox{\scriptsize exp}}$ and then
$\Delta^{\mbox{\scriptsize stat}} M_{t(n)}$ and
$\Delta^{\mbox{\scriptsize sys I,II}} M_{t(n)}$. Since in our
model the pattern of the dependence on $n$ is the common one in
the LC and LHC cases, the difference between these cases appears
in the scales of the errors only. This allows us to present the
results on Figs.~\ref{Fig4}--\ref{Fig6} with adding new scales.
The numerical results are presented in Table~\ref{T2}. It is
interesting to note that $\Delta^{\mbox{\scriptsize sys I}}
M_{t(1)}$ turns out smaller than $\Delta^{\mbox{\scriptsize sys}}
M_{t}$ obtained in Ref.~\cite{Chekanov1} in the framework of the
direct reconstruction of events. Nevertheless this does not mean
an inconsistency. Really, we equate $\Delta^{\mbox{\scriptsize
sys}} M_{t}$ of \cite{Chekanov1} to $\Delta q$ at $q \simeq M_t$,
but the dominant contributions to $\Delta^{\mbox{\scriptsize sys
I}} \langle q \rangle_{\mbox{\scriptsize exp}}$ are formed at
strictly smaller $q$ than $M_t$, which is obvious from formulas
(\ref{Eq12}) and (\ref{Eq13}). This effect diminishes
$\Delta^{\mbox{\scriptsize sys I}} M_{t(1)}$ compared to
$\Delta^{\mbox{\scriptsize sys}} M_{t}$ of \cite{Chekanov1}.

\begin{table}[t]
  \centering
\begin{tabular}{c c c c c c }
\hline
& & & & & \\[-3mm]
  n & $\Delta^{\mbox{\scriptsize stat}} M_{t(n)}$
  & $\Delta^{\mbox{\scriptsize sys I} } M_{t(n)}$
  & $\Delta^{\mbox{\scriptsize sys II}} M_{t(n)}$
  & $\Delta^{\mbox{\scriptsize sys}   } M_{t(n)}$
  & $\Delta \! M_{t(n)}$ \\[1mm] \hline
& & & & & \\[-3mm]
 1  &0.80        & 0.17        & 0.19        & 0.26
                                                   & 0.84 \\
 5  &0.24        & 0.04        & 0.13 (0.11) & 0.13 (0.12)
                                                   & 0.27 \\
 10 &0.17        & 0.02        & 0.13 (0.09) & 0.13 (0.09)
                                                   & 0.27 (0.20) \\
 15 &0.16        & 0.01        & 0.15 (0.08) & 0.15 (0.08)
                                                   & 0.22 (0.17) \\
 20 &0.16 (0.15) & 0.01 (0.00) & 0.21 (0.07) & 0.21 (0.07)
                                                   & 0.26 (0.16) \\
 30 &0.23 (0.15) & 0.01        & 0.41 (0.07) & 0.41 (0.07)
                                                   & 0.46 (0.16) \\
 [0.5mm] \hline
\end{tabular}
\caption{\small The same that in Table~\ref{T1} in the LC
case.}\label{T2}
\end{table}

\section{Theoretical uncertainty\label{theor}}

The analysis of the previous sections shows that the experimental
accuracy of the $M_t$ determination can be considerably improved
by the transiting to the high degrees of the moments. So, for the
achieving of the eventual high accuracy a theoretical uncertainty
becomes more and more crucial. In this connection it is important
to understand whether the theoretical error in the $M_t$
determination can be made smaller than the experimental error at
the high degrees of the moments. If it will be possible then the
theoretical error will not spoil the expected accuracy. Below we
discuss this question in somewhat qualitative manner since the
possibility of the solution first of all is important.

Let us begin with the note that the origin of the theoretical
uncertainty in the $M_t$ determination is connected with the
uncertainty of the calculation of the theoretical moment $\langle
q^n \rangle$ in equation (\ref{Eq5}). Further, at the determining
of $\Delta \! M_{t(n)}$ the corresponding error
$\Delta^{\mbox{\scriptsize th}} \langle q^n \rangle$ is to be
added (in quadratures) to $\Delta \langle q^n
\rangle_{\mbox{\scriptsize exp}}$ in formula (\ref{Eq6}). So the
problem is reduced to the question about a possibility of carrying
out the calculations so precisely that to keep the error
$\Delta^{\mbox{\scriptsize th}} \langle q^n \rangle$ smaller than
$\Delta \langle q^n \rangle_{\mbox{\scriptsize exp}}$.

In practice it is convenient to compare relative errors like
$\Delta \langle q^n \rangle / \langle q^n \rangle$ instead of the
proper errors $\Delta \langle q^n \rangle$. Fortunately the
experimental relative error $\Delta \langle q^n
\rangle_{\mbox{\scriptsize exp}}/\langle q^n
\rangle_{\mbox{\scriptsize exp}}$ is growing at the transiting to
the high degrees of the moments; its behavior is similar to that
represented in Fig.~\ref{Fig2}. In particular, at transiting from
$n=1$ to $n=15$ the $\Delta \langle q^n \rangle_{\mbox{\scriptsize
exp}}/ \langle q^n \rangle_{\mbox{\scriptsize exp}}$ increases
from 1.8\% to 5.2\%(4.5\%) in the LHC case, and from 0.7\% to
2.4\%(1.9\%) in the LC case. So with increasing $n$ the
requirement for the theoretical relative error
$\Delta^{\mbox{\scriptsize th}} \langle q^n \rangle/ \langle q^n
\rangle$ is weakening.

Generally a theoretical error arises from a parametric uncertainty
and an intrinsic uncertainty of the itself calculation. The
parametric uncertainty originates mainly from the parameters that
are worse known. In the given case they are the widths of the $W$
boson and of the top quark. The analysis of
Ref.~\cite{Kharchilava} shows that the uncertainties in these
parameters practically do not affect the first moment. Moreover,
even the switching-off of the widths gives a negligible effect. In
the 15-th moment, the varying of $\Gamma_W$ within the
experimental error $\Delta \Gamma_W = 0.04$~GeV results in $\Delta
\langle q^{15} \rangle/ \langle q^{15} \rangle = 0.09\%(0.06\%)$,
which is insignificant, as well. Unfortunately we cannot estimate
the variance of the moments with varying the width of the top
quark, since from the very beginning we use for the top quarks the
narrow width approximation. Nevertheless by basing on the results
of Ref.~\cite{Kharchilava} we expect a negligible variance of the
moments in this case too. This is corroborated by the lack of
reasons leading to appreciably greater sensitivity of the moments
with respect to the width of the top quarks than to the width of
the $W$ boson.

It should be mentioned, however, that the complete switching-off
of the widths can vary noticeably the high-degree moments. Thus,
the setting $\Gamma_W\!=\!0$ implies a shift of $\langle q^{15}
\rangle$ on 4.6\%(3.2\%) which can be compared with the above
estimations for the experimental relative errors. This means that
the calculation of the high-degree moments must be carried out
with the taking into consideration of the realistic values of the
widths. The latter requirement, of course, is unnecessary for the
estimation of the errors only, the case of the investigation of
the present article.

Now let us consider the errors of the itself calculation. First we
note that all the processes in (\ref{Eq1}) go far above the
thresholds of the production of unstable particles, the $W$ bosons
and the top quarks. Therefore their production and decay can be
described by the standard methods \cite{4f}, namely with the Dyson
resummation in the leading order calculation and in the pole
approximation at calculating the perturbation-theory
corrections.\footnote{As variants, one can exploit the method of
an effective field theory for calculating the resonant processes
\cite{Chapovsky} or the modified perturbation theory based on
distribution theory \cite{F,Nekrasov}.} Thus, the problem is
reduced to the estimation of the order of the perturbation theory,
which is necessary for satisfying the required precision of the
calculation.

Further we note that the corrections to the moments are to be
calculated by way of calculating the corrections to the
distribution $F(q)$. For kinematic reasons the behavior of the
latter corrections in the basic features should follow the
behavior of the distribution. Namely, $\Delta^{\mbox{\scriptsize
th}} F(q)$ must vanish at the ends of the kinematic region because
of the vanishing phase volume. Furthermore,
$\Delta^{\mbox{\scriptsize th}} F(q)$ must almost vanish on the
tail at large $q$ due to the smallness of the width of the $W$
boson. (Recall that in the limit $\Gamma_W = 0$ the distribution
is completely suppressed on the tail by reason of kinematics.) So
$\Delta^{\mbox{\scriptsize th}} F(q)$ must be precisely known
mainly in the middle of the kinematic region but not near its ends
including the tail. At transiting to the high-degree moments this
condition is maintained. Moreover, in some sense it becomes even
more strong. Really, at low $q$ the contributions to the moments
are additionally suppressed by the factor $q^n$. At large $q$, in
the case of the effective moments, the contributions are
completely suppressed by the cutoff $\Lambda_n$. In addition, the
larger $n$ the more the distant between the cutoff and the $M_t$,
the right boundary of the actual range of kinematic variable
(since $\Lambda_n \to \Lambda$ from the right as $n \to \infty$).
In particular, $\Lambda_{1} = 171$~GeV but $\Lambda_{15} =
160$~GeV which is distant form $M_t$ by 15~GeV. The mentioned
feature is valuable for our consideration as the cut-off of the
ends of the region of the kinematic variable implies a suppression
of large logarithms that can arise near the ends at calculating
the perturbation-theory corrections. In the final analysis this
allows us to use the naive counting method for estimating the
corrections to the moments.\footnote{It should be emphasized that
we discuss here the corrections to the $t \to b$ transition but
not to the $b$-quark fragmentation, including the perturbative
fragmentation. The latter process is described by the convolution
of the cross-section with the fragmentation function, and this
operation is to be fulfilled in the framework of MC event
generators.}

With this in mind we farther use a rather rough approach which is
based on a comparison between the corrections to the moments and
to the width of the top quark. (Notice that the width actually is
the zero moment accurate to the normalization.) The key reason of
the approach is the observation that the integrals for the moments
and the width, and for the corrections to the moments and the
width, accumulate the contributions mainly from the middle region
of the kinematic variable. So, supposing that in this region the
correction to the distribution $\Delta^{\mbox{\scriptsize th}}
F(q)$ varies weakly in the units of $F(q)$, one can expect that
the corrections to the moments and to the width should be close to
each other in the relative units. By the closeness we admit here a
factor of order of several units. It is worth mentioning that even
in the case of the experimental errors, which depend strongly on
the shape of the distribution, the relative errors at $n=1$ and
$n=15$ differ from each other by a factor of 2.5--3.5 only.

As we know, the electroweak one-loop correction to the top quark
width amounts approximately 2\%. The QCD one-loop correction is
near of 10\%, while the two-loop one is near of 2\%. (See
\cite{top} and the references therein.) The comparison of these
values with the above estimations for the experimental relative
errors demonstrates that the one-loop electroweak and two-loop QCD
corrections are enough to remain within the required limits (in
both cases, LHC and LC). It should be noted that the mentioned
corrections to the distribution can be certainly calculated since
the corrections to the width have been calculated. Finally we note
also that only the direct calculations of these corrections can
explicitly solve the problem of the theoretical errors to the
moments.

The mentioned calculation, however, would not yet entirely close
the problem of the theoretical error of the $M_t$ determination
because of the problem of nonperturbative nature caused by a
renormalon contribution. Below for the completeness we only
briefly consider this problem as its solution is known, at least
in a conceptual respect. The problem actually is connected with
the kind of the mass which is to be determined through an
experimental measurement. In fact there are different masses, but
only a Lagrangian mass is ultimately valuable since only the
Lagrangian mass can be constrained with other fundamental
parameters of the theory. The important representatives of the
Lagrangian mass are the pole and $\overline {\mbox{MS}}$ masses.
The directly measurable one is the pole mass, which is determined
by kinematics. Correspondingly, the currently used algorithms of
extracting $M_t$ from the data are turned to the pole mass.
However, because of the renormalon contribution the pole mass
determination faces an extra uncertainty of order of
$O(\Lambda_{QCD})$ \cite{Bigi}. Numerically it can amount hundreds
of MeVs.

The above mentioned difficulty can be bypassed in the framework of
the following algorithm (below we state one of its possible
variants) \cite{Bigi}. First, all theoretical calculations are to
be fulfilled in the terms of the pole mass. Then the value of the
pole mass is to be determined from the matching with data.
Remember, at this stage the result includes the renormalon
contribution. Further, by means of the well-known formula relating
the pole mass with the $\overline{\mbox{MS}}$ mass (see
Ref.~\cite{top}, for example), the $\overline{\mbox{MS}}$ mass is
determined. At this step the result gets again the renormalon
contribution but, as is declared, it cancels the previous one. (So
the inaccuracy in the relation between the pole and
$\overline{\mbox{MS}}$ masses is charged to the pole mass.) Direct
calculations in certain examples \cite{threshold,1S} demonstrate
effectiveness of the above algorithm.

So, the problem is initially stated as though for the pole mass
determination, but at the final stage the $\overline{\mbox{MS}}$
mass is determined. This allows one to avoid a theoretical
systematic uncertainty of order of $O(\Lambda_{QCD})$ caused by
the renormalon contribution. Returning to our outcomes, we see
that the theoretical error of the top mass determination can be
really made smaller than the experimental error.

\section{Discussion\label{Discussion}}

The major result of this article is the detection of the effect of
decreasing of statistical and systematic errors of the top quark
mass measured from $M_{b\ell}$ distribution, when applying the
technique of the moments and proceeding to the high degrees of the
moments. The optimal value of the degree minimizing the errors is
found near $n = 15$.

\begin{figure}
\hbox{ 
       \epsfxsize=210pt \epsfbox{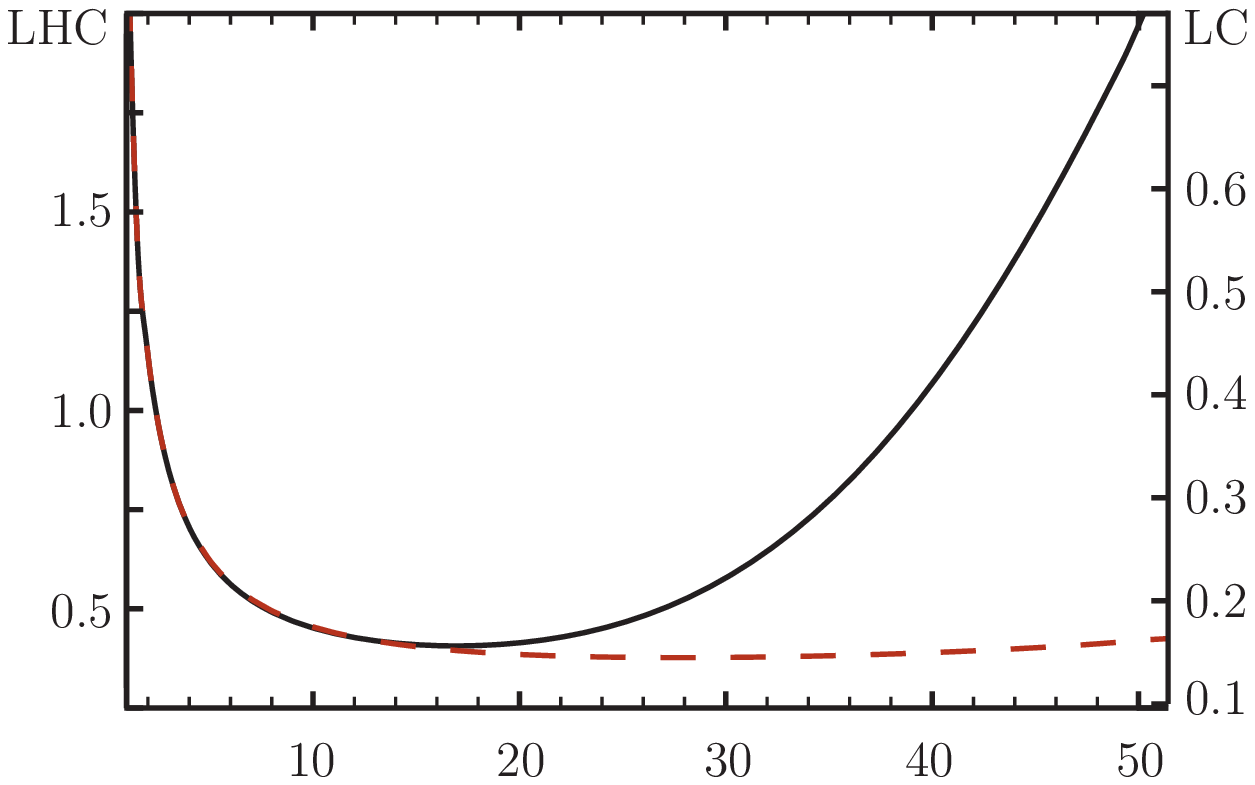}}
\caption{\small The statistical error $\Delta^{\mbox{\scriptsize
stat}} M_{t(n)}$ depending on $n$. The dashed curve represents the
results obtained by the method of the effective moments. The left
and right vertical axes scale in GeV the results for LHC and LC
cases, respectively.}\label{Fig4}
\end{figure}

\begin{figure}
\hbox{ 
       \epsfxsize=210pt \epsfbox{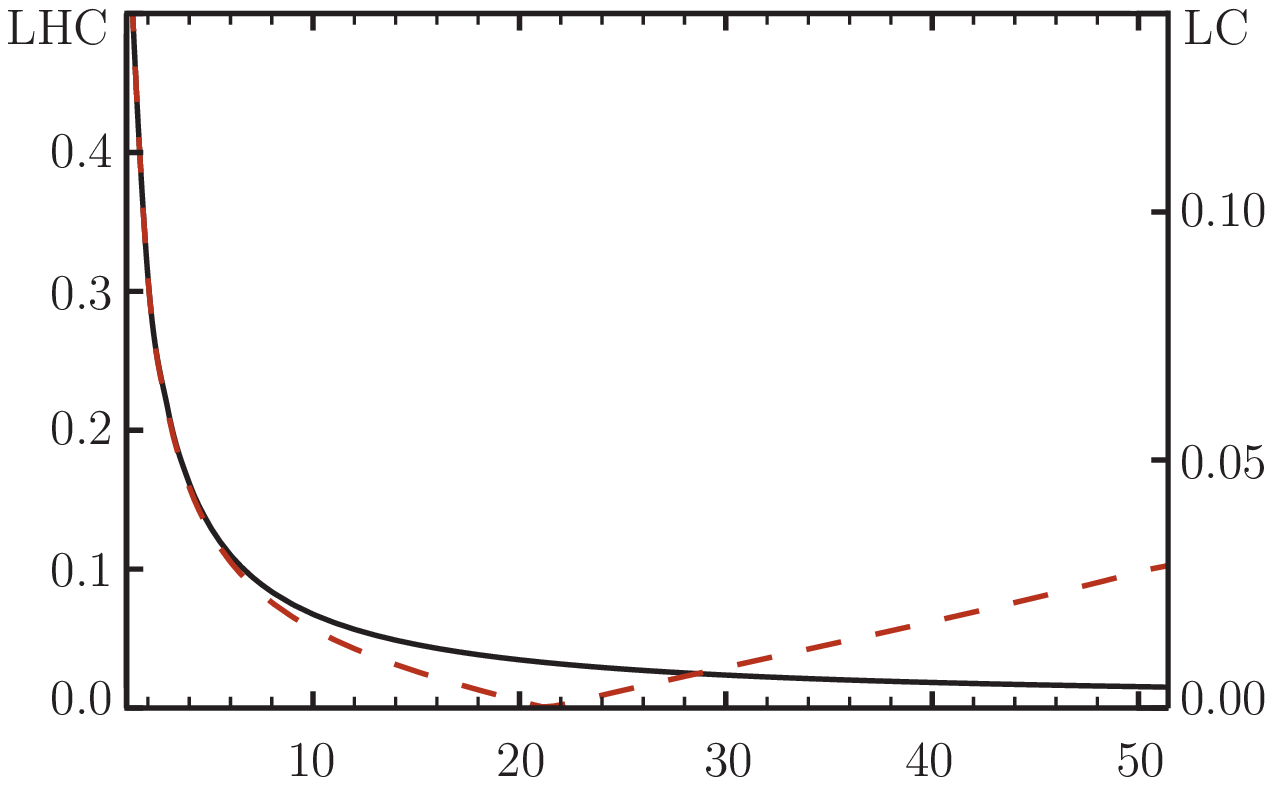}}
\caption{\small The same that on Fig.~\ref{Fig4} for
$\Delta^{\mbox{\scriptsize sys I}} M_{t(n)}$.}\label{Fig5}
\end{figure}

\begin{figure}
\hbox{ 
       \epsfxsize=210pt \epsfbox{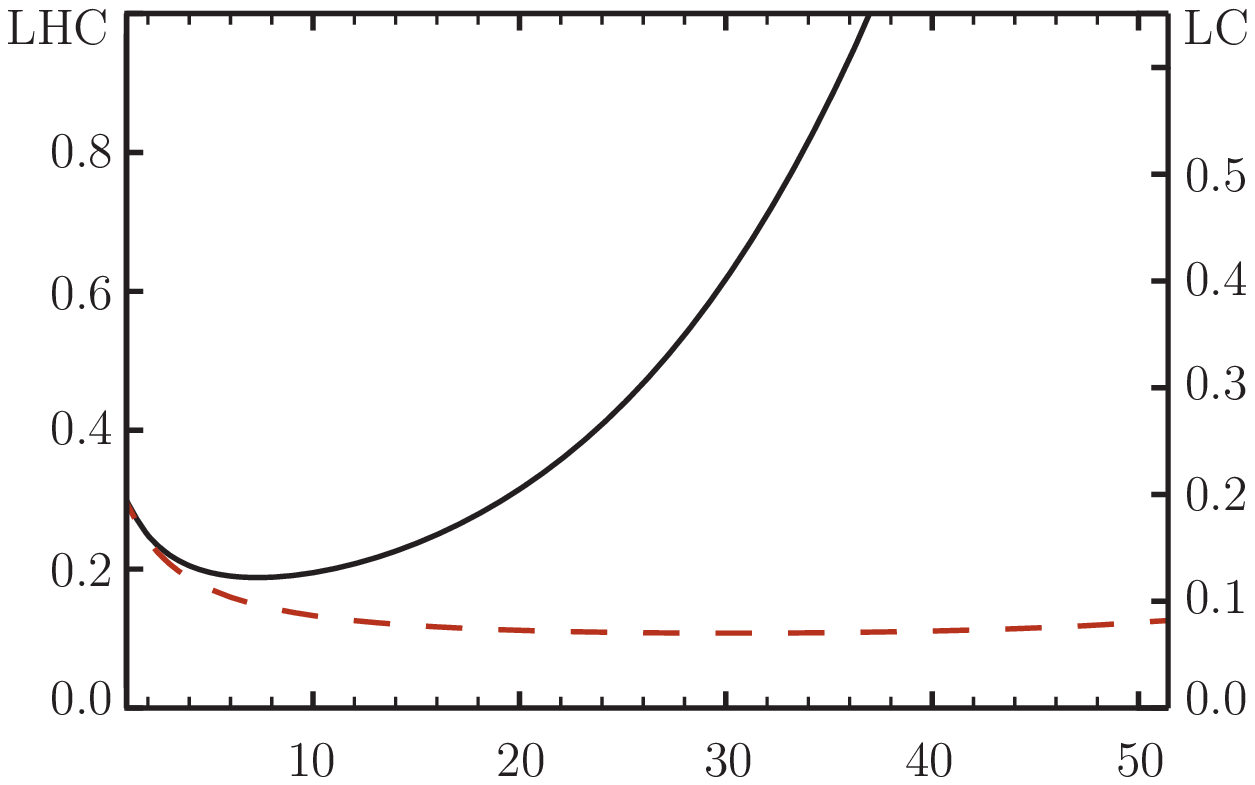}}
\caption{\small The same that on Fig.~\ref{Fig4} for
$\Delta^{\mbox{\scriptsize sys II}} M_{t(n)}$.}\label{Fig6}
\end{figure}

For the determining of the errors we have attracted a simple
enough model. Its parameters in the LHC case have been fixed on
the basis of the results obtained earlier \cite{Kharchilava} by
the MC modelling method. As applied to LC the parameters have been
fixed mainly by indirect methods. Knowing the parameters and the
dependence on the degree $n$ of the moments, we have estimated the
errors at varied $n$ and have found the optimal value of $n$,
minimizing the errors. The optimal value $n=15$ is clearly visible
in the framework of the basic method of calculating the moments.
The applying of the technique of the effective moments diminishes
the errors at $n=15$ by 10-20\%, but at the further increasing of
$n$ the results practically do not vary (see
Figs.~\ref{Fig4}--\ref{Fig6} and Tables~\ref{T1}--\ref{T2}).

At the optimal value $n = 15$ the total error $\Delta M_t$ is
found close to 500 MeV in the LHC case, and close to 200 MeV in
the LC case. In the LHC case the above accuracy more than in twice
exceeds the accuracy obtained by the other methods \cite{top}
including the original method of Ref.~\cite{Kharchilava}. In the
LC case the estimated accuracy of the $M_t$ determination is close
to that expected at scanning the $t \bar t$ production threshold
\cite{threshold,1S}.

In conclusion it should be mentioned, once again, that at the
intermediate stage of the analysis we have introduced
simplifications allowing to minimize calculations. However at the
final stage all estimations have been made on the basis of
realistic values of the parameters. This peculiarity should not
reduce the legitimacy of the detected behavior of the errors and,
moreover, of their rounded estimations. Nevertheless the
quantitative outcomes could be improved by further calculations
based on the direct applying of a proper MC event generator.

\newpage

\twocolumn

\onecolumn

\end{document}